\documentclass[12pt]{elsarticle}
\usepackage{amssymb, amsmath}
\usepackage[utf8]{inputenc}
\usepackage[english]{babel}
\usepackage{mathtools}
\usepackage{soul} 
\usepackage{amsthm}

\newtheorem{prop}{Proposition}

\usepackage{graphicx}   
\usepackage{multicol}
\usepackage{caption}
\usepackage{subcaption}

\begin{document}

\begin{frontmatter}



\title{Analysis of potential flow networks: Variations in transport time with \textit{discrete}, \textit{continuous}, and \textit{selfish} operation}


\author[inst1,inst2]{Varghese Kurian}

\author[inst1,inst3]{Sridharakumar Narasimhan}%

\affiliation[inst1]{organization={Department of Chemical Engineering, Indian Institute of Technology Madras},
            city={Chennai},
            postcode={600036}, 
            state={TN},
            country={India}}
\affiliation[inst2]{organization={Department of Chemical and Biomolecular Engineering, University of Delaware},
            city={Newark},
            postcode={19716}, 
            state={DE},
            country={USA}}            
\affiliation[inst3]{organization={Robert Bosch Center for Data Science and Artifical Intelligence, Indian Institute of Technology Madras},
            city={Chennai},
            postcode={600036}, 
            state={TN},
            country={India}}

\begin{abstract}
In potential flow networks,  the equilibrium flow rates are usually not proportional to the demands and flow control elements are required to regulate the flow. The control elements can broadly be classified into two types – discrete and continuous. Discrete control elements can have only two operational states: fully open or fully closed. On the other hand, continuous control elements may be operated in any intermediate position in addition to the fully open and fully closed states. Naturally, with their increased flexibility, continuous control elements can provide better network performance, but \emph{to what extent?} 

 We consider  a class of branched networks with a single source and multiple sinks. The potential drop across edges ($\Delta H$)  is assumed to be proportional to  the $n^{th}$ power of flow rate ($Q$), i.e., $\Delta H=kQ^n$ , ($n\geq1$). We define \textbf{R} as the ratio of minimal operational times required to transport a given quantum of material with either type of control element and   show that $1\leq \textbf{R}\leq m^{\left(1-1/n\right)}$, where $m$ is the maximum depth of the network. The results point to the role of network topology in the variations in operational time. {Further analysis reveals that the selfish operation of a network with continuous control valves has the same bounds on the price of anarchy.}
\end{abstract}




\end{frontmatter}


\section{Introduction} \label{sec:Introduction}
    
In \textit{potential flow networks}, flow through the edges are driven by the potential difference across them. 
The equilibrium flow in these networks can be computed by solving the balance equations for the nodes and the potential loss equations for the edges - sometimes referred to as the \textit{optimal flow problem} \cite{rockafellar1984network}. In many systems (water networks, biological transport, gas networks; see \cite{hendrickson1984common,
raghunathan2013global,miguel2018general,ronellenfitsch2016global,klimm2022robustness} for examples of potential flow networks), the desired flow rates in the network may be different from the equilibrium flow rates. In such situations, it is common for an external/internal agent to alter the potential difference or the network resistance across  the arc(s) in the network to attain the desired flow rates (valve/pump operations, vasodilation/constriction, flow regulation). A time-varying profile of network alterations is referred to as the schedule for operation or control of the network. For an agent modifying the network resistance, the capability to manipulate the network resistance can be of two types - (i) open or close the edges to start or stop the flow through them (ON/OFF or discrete control) or (ii) continuously modify the network resistance and thereby keep the edges partially open (continuous control). The flexibility and cost associated with the control system could vary with the type of actuation. The primary objective of this paper is to quantify the variation in performance of the network with the type of control available. An extension of the results also highlights the differences between the optimal and \textit{selfish} operation of a network with continuous control valves. The results are general and  are applicable to systems in which the flow rates across the edges grow  sub-linearly with the potential difference. Though there has been several recent works on complex networks, including those on controllability \cite{liu2016control}, the authors have not come across any prior work discussing the variation in network performance with the type of actuation. 
   
 \section{Problem statement}
 \subsection{System description} \label{sec:sys_description}
\begin{figure} [ht]
    \begin{center}
	\includegraphics[width=0.7\linewidth]{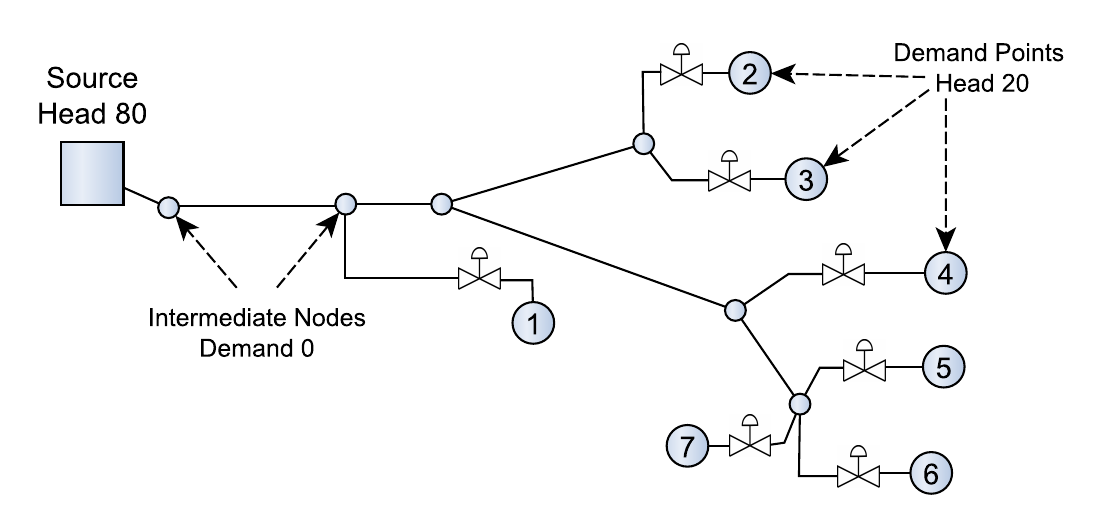}
	\caption{Schematic of a water network with a single source and seven demand nodes. Here the demand points are at the same head and the source is maintained at a higher head. The resistance in the links immediately upstream of the demand nodes can be manipulated to control the flow through them. }
	\label{fig:sample_network1}
	\end{center}
\end{figure}

The system addressed here consists of a  single source and multiple sinks as shown in Figure \ref{fig:sample_network1}. The sinks - also referred to as demand points/demand nodes - are all assumed to be at the same potential and the source is maintained at a higher, but constant potential.  The network is assumed to be a branched network (with no loops) and the resistance of every edge is assumed to be non-negative.  Although we specialize the subsequent analysis to water networks, an important class of engineered networks, purely for ease of exposition,  the analysis is quite general. 

The potential drop  across the edges $(\Delta H)$ are assumed to be of the form
\begin{align}
\Delta H= k Q^n,~ n\geq 1 \label{eq1}
\end{align}
\noindent where  $Q$ is the flow rate through the edge. In the case of water flow, $n$ is typically assumed to be 1.85 and the potential is usually referred to as head \cite{wright2015control}. In gas transportation networks $1.8\leq n\leq2$ \cite{shiono2016optimal}.
There is a predefined quantity of material (water here) that has to be transported from the source to the sinks and the demand may vary between sinks. Further, it is assumed that non-zero demands exist only at the leaf nodes and the resistance of the edges leading to any sink (terminal edges) can be manipulated (controlled) to meet the demands correctly.   

\subsection{Problem description} 
 The problem considered here involves  transporting a given amount of water from the source to sinks in the minimum time. As the equilibrium flow rates may not be proportional to the demands at the sinks, one has to suitably operate the manipulated variables (valves) to ensure that the correct amount of water is transported. Naturally, the minimum time required to meet the demand can vary with the type of control. The available choices for the manipulated elements in discrete control is only a subset of the same in continuous control. Therefore, the minimum operational time using discrete control has to be greater than or equal to the minimum operational time using continuous control.  This leads to the central question addressed in this work - $\mathcal{P}1:~$\textit{For a given network and demand, to what extent can the operational times vary with the type of control available?} Let the ratio of time required in discrete control to that in continuous control  be defined as \textbf{R}. This paper gives an upper bound on \textbf{R} for the class of branched networks described earlier, that is  general and independent of the demands or edge resistances. 
 
 Further, this work  extends the discussion to the 
 cost of selfish operation. If the control of the network elements is given to the respective demand nodes rather than managed centrally, the operational objective of the agents would be to minimize the time for completing their respective individual demand rather than the total network demand. The resultant operation need not be  optimal for the network. In this context, a second question that is relevant is -  $\mathcal{P}2:~$\textit{For a network with continuous control elements, to what extent can the operational times vary with the agent - centralised or decentralised - managing the system?}  Later sections of this paper shows that the the bounds for this second problem $(\mathcal{P}2)$ is same as that of the initial problem $(\mathcal{P}1)$.

    \section{ Results} \label{sec:Results}

\subsection{A trivial bound for \textbf{R} }

\textit{Lower bound:} The set of valve configurations available for discrete valves is a subset of the same for continuous control valves. Any system state configured using discrete valves can also be obtained using continuous control valves.  Hence, the lower bound on the ratio of supply times is unity.  

\textit{Upper bound:} Let $t_{cv}$ be the time for which a network has to be operated to meet the demand using continuous control valves. A simple schedule for supply using discrete valves would be to supply water to one demand point at a time. Using  Proposition \ref{prop_1} it can be shown that in this simple schedule,  each demand point will be supplied for a time interval less than or equal to $t_{cv}$. Altogether, the total time required for supply using discrete valves would be less than or equal to $\left| T \right| \times t_{cv} $ where $\left| T \right|$ is the number of demand points in the network. This gives a somewhat trivial upper bound for the ratio of supply times: $\frac{\left| T \right| \times t_{cv}}{t_{cv}}$. Hence we can write, 
\begin{align}
    1 \leq \mathbf{R}\leq \left| T \right|, ~~~ \left| T \right| \text{being the number of demand nodes} \label{eqn:trivial bound}
\end{align}

However, this bound is not tight and the upper bound ($\left| T \right|$) can be quite misguiding. A  non-trivial, but tight bound for the same case is presented in the later sections. 

While computing a bound on $\mathbf{R}$, by definition, one would have to consider all possible discrete schedules and continuous schedules over all possible networks, topologies and demands. The following results allow us to obtain this bound in a systematic manner by identifying key   properties that maximize $\mathbf{R}$. The derivations for these results are given in Appendix \ref{sec:results_proof}. 

\subsection{The supremum of \textbf{R} is given by a scenario where the 
continuous control valves  are maintained at a constant setting throughout the operation. 
} \label{theorem_1}

\noindent In general, when operating the network using continuous control valves, the valve positions and hence the flow rates through the network can change over time. However, the above result allows us to restrict our search space to operational schedules with constant valve positions.
Here, the  settings of the continuous control valves would be such that the flow rates received by demand points are proportional to their respective total demand. 

\subsection{The supremum of \textbf{R} is given by a class of networks with all edges, but the ones belonging to a `main path', being of zero resistance}  \label{theorem_2}
\begin{figure} [htpb]
    \begin{center}
	\includegraphics[width=0.7\linewidth]{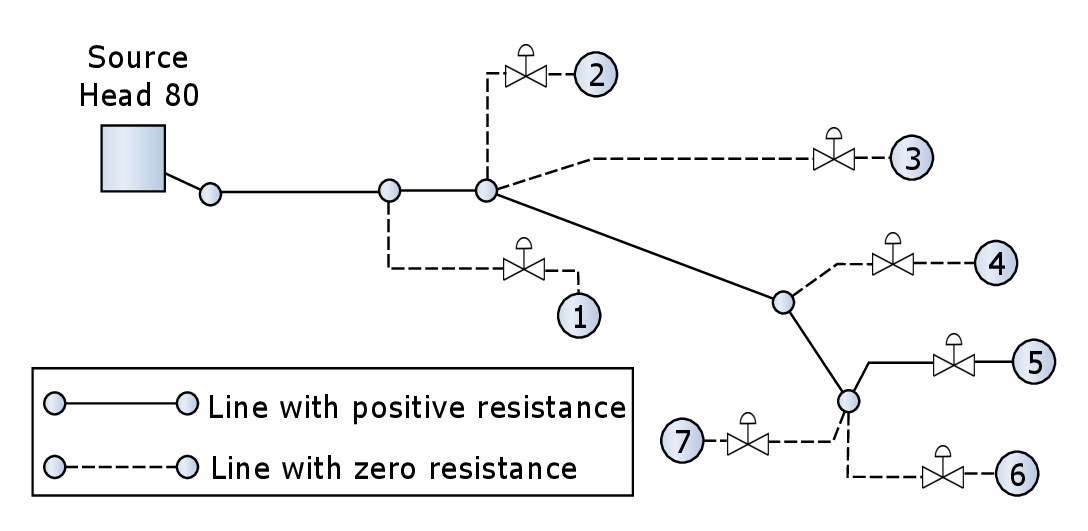}
	\caption{A network belonging to the class $\mathcal{C}$. Here only the line from source to node $5$ have positive resistance when fully open. Remaining nodes are connected to this main line with an edge of zero resistance. }
	\label{fig:theorem2}
	\end{center}
\end{figure}

\noindent The supremeum of \textbf{R} is realized in a particular class of networks. Here, the main line -  connecting the source to one of the leaf nodes - would have positive resistance. All other demand points would be connected directly to this main line with an edge of zero resistance. We denote this class of networks as $\mathcal{C}$. An example of one such network is given in Figure  \ref{fig:theorem2}.


\noindent  

\subsection{For a network with the characteristics mentioned in Section \ref{sec:sys_description}, a tight bound on the ratio of operational times is given by:} \label{theorem_3}
\begin{align}
    1\leq  \mathbf{R}  \leq m^{1-1/n}
\end{align}
where $m$ is the maximum depth of the network.

\section{Discussion}
\subsection{At $m=2$, the upper bound equals the expression given in \cite{Velmurugan2023Continuous}} \label{sec:verification-preliminary}
Earlier, in \cite{Velmurugan2023Continuous}, we presented a bound on operational times for a network with three pipes. A closed form expression for \textbf{R} could be derived and the optimum identified analytically.  At $m=2$, the upper bound given in Result \ref{theorem_3} turns out to be the same as the one presented in \cite{Velmurugan2023Continuous} for three pipe networks. It may be noted that the only part common for both derivations was the first step (Result \ref{theorem_1}). Hence, this serves as a preliminary verification of the current result. 

\subsection{Comparing the tight bound with the trivial bound} \label{sec:comparison-tight, trivial bounds}
Two aspects of the tight bound given by Result \ref{theorem_3} are notable, particularly in light of the trivial bound given by Equation \ref{eqn:trivial bound}. 

(i) Result \ref{theorem_3} states that the bound on \textbf{R} is dependent on the depth of the network. A direct consequence of this is the role of network topology in the variations in supply time, a factor that cannot be inferred from common knowledge or Equation \ref{eqn:trivial bound}. Two networks having the same number of demand points, but different depths, behave differently. When $m=1$,   i.e., the network has a star topology (all demand nodes directly connected to the source),  $\mathbf{R}=1$. In such networks, every demand node can be independently controlled and hence, discrete and continuous operations are equivalent. However, for the same number of demand nodes, topologies with larger depth  will require more time to supply the same amount of water when operated with discrete valves.  

(ii) The tight bound on \textbf{R} increases sub-linearly with the maximum depth of the tree. 
This second aspect implies that the variation in supply times scales with the size of the network at a much slower rate compared with Equation \ref{eqn:trivial bound}. 
 
\subsection{Topology and inequity}
$\mathbf{R}$ is  an upper bound on the ratio of times taken for discrete operation as compared to continuous operation. It can be used to understand and quantify inequity in supply. Under distress conditions or due to other resource constraints, it is common practice to operate the network for a limited amount of time thus limiting the amount of water supplied. The resultant supply is said to be inequitable when the fraction of the demand met varies substantially between the nodes. 

Network topologies of the type shown in Figure \ref{fig:theorem2}, in general, can lead to high inequity. In these systems, the demand point closest to the source has an easier access to water and the farther nodes would receive supply only after the earlier valves are closed. 
This implies that the  demand nodes away from the source are at a disadvantage in the absence of good operational policies. 
The fact that the value of $\mathbf{R}$ can be  high  for these network topologies makes it even worse for the farther nodes if available time is limited. From the perspective of inequity, it would appear that  trees with small depth ($m$) are better than unbalanced trees if the available control is of discrete type.  A star topology is ideal ($m=1$), but might be expensive to build for most practical applications. On the other hand, the maximum depth of a  tree network with $l$ nodes is $l-1$ which happens to be a chain. , the network is essentially a  chain, i.e., $e_i=(v_i,v_{i+1}), i=1,..l-1$. It is to be expected that the network is difficult to control, viz.,when all valves are open,  the node closes to the source will receive disproportionally more water than the farthest node.  



\subsection{Decreasing network resistance can increase the total supply time} \label{sec:decrease_resistance_increase_time}
Braess's paradox is \textit{the counter-intuitive phenomenon that removing arcs from a network can improve the cost of selfish routing} \cite{roughgarden2003price} (for examples of Braess's paradox in physical systems, see \cite{case2019braess,bittihn2018braess,
ma2019airway}).  A similar phenomenon is observed here when the total time required for transportation is considered as the \textit{cost}. 

Consider again the network in Figure \ref{fig:theorem2}. Let the network be equipped with continuous control valves. Two cases are studied. In the first case, all valves are operated by an external agent with the objective of minimizing the total operational time of the network. In the second case, the control is given to individual demand points, each having the selfish objective of minimizing the time required to meet their respective demands.

Case (i): Centralised control of the network. Let the system be operated under a single configuration of continuous control valves for time $t_{cv}$. Assume that the resistance offered by all valves are positive, except the valve leading to Node 5. The valve leading to Node 5 is kept fully open (zero resistance).   Let the quantum of water supplied in time $t_{cv}$ be equal to the demand of the network.

Case (ii): Decentralised control of the network. Here each demand node has the control of the valve leading to 
it and these agents operate with the selfish objective of minimizing the time for meeting their respective demand. To minimize the time, every demand point starts with their valves fully open. When the demand of a node is met, the particular valve is closed. As the resistance offered by all edges originating from the main line are zero, the node closest to the source (Node 1) would receive all water initially. The second node receives water after the demand of the first node is met. Likewise, during the entire time the network is operated, only one node receives water at a time. From Appendix \ref{sec:proof_result2} it can be inferred that the time taken for this strategy is $\mathbf{R}\times t_{cv} $ with $\mathbf{R}>1$. 

Now, in Case (ii), if the terminal edges of the network had a higher resistance - of the same extent as the resistance offered by the valves in Case (i) - the entire demand could have been met in time $t_{cv}$, even if the control was decentralised. This shows that in a network following a decentralised (selfish) operation, reducing the resistance in certain edges can increase the total operational time - much like Braess's paradox. 
Note that the paradox  quoted here refers to the time varying behaviour of a \textit{dynamic} network, unlike the \textit{static} network considered in \cite{calvert1993braess}.



\subsection{Bound on the ratio of supply times is also the bound on the \textit{price of anarchy} (PoA) for a system with continuous control valves}

In Section \ref{sec:decrease_resistance_increase_time}, the first case had an external agent operating the valve resistances. This would have been the optimal operational strategy for the network considering the overall goal of minimizing the supply time. The second case followed a selfish operation by individual demand points. 
Define the ratio of operational times in the selfish operation to that of the optimal operation is defined as the \textit{price of anarchy} (more information on PoA can be found in \cite{roughgarden2003price})


Now, consider a network with continuous control valves. Let $t_{cv}$ be the time to complete the supply for an agent having the objective of minimizing the network operational time. Let there be a second scenario where the control of individual valves is with the respective demand node. The operational objective of each node is  to minimize the time for fulfilling their individual demand. The ratio of time required for selfish operation to that of optimal operation gives the price of anarchy (PoA) for this dynamical network.  

In the decentralised control scheme, the selfish operational strategy of each node would be to keep their valve fully open from the beginning, until their demand is completely met. This is the schedule $\mathcal{S}$ described in Section \ref{sec:proof_result2}. Further, Section \ref{sec:proof_result2} identified the class of networks (class $\mathcal{C}$; an example given in Figure \ref{fig:theorem2}) for which Schedule $\mathcal{S}$ requires the longest operational time. Hence, the ratio of selfish to optimal operational time of the networks in class $\mathcal{C}$ gives an upper bound for the PoA. The expression derived in Section \ref{sec:An_expression_for_R*},  $\mathbf{R^*} = m^{1-1/n}$, therefore, is also a bound on the price of anarchy. 

We often consider hierarchical trees as optimal networks based on notions of dissipation minimization and adaptability \cite{
ronellenfitsch2016global, 
kou2014optimal, 
liao2021narrowing}. On these metrics, unbalanced trees with resistances decreasing near the leaf nodes  usually perform poorly. The networks in class $\mathcal{C}$ fall in this category. It is now clear that these `poor designs'  also perform poorly in the sense of PoA, if  the transportation time is considered as the cost.

\section*{Acknowledgments}
This work was partially supported by the Department of Science and Technology, Govt. of India under the Water Technology Initiative (DST\slash TM\slash WTI \slash WIC \slash 2K17\slash 82(G)- WATER-IC for SUTRAM of EASY WATER).




 \bibliographystyle{elsarticle-num} 
\bibliography{PhysicaA_R0}

\newpage

\appendix    

\noindent \large{\textbf{Appendices}}

\normalsize{}
\renewcommand{\appendixname}{}
\section{Preliminary results}

\begin{figure} [htpb]
    \begin{center}
	\includegraphics[width=0.7\linewidth]{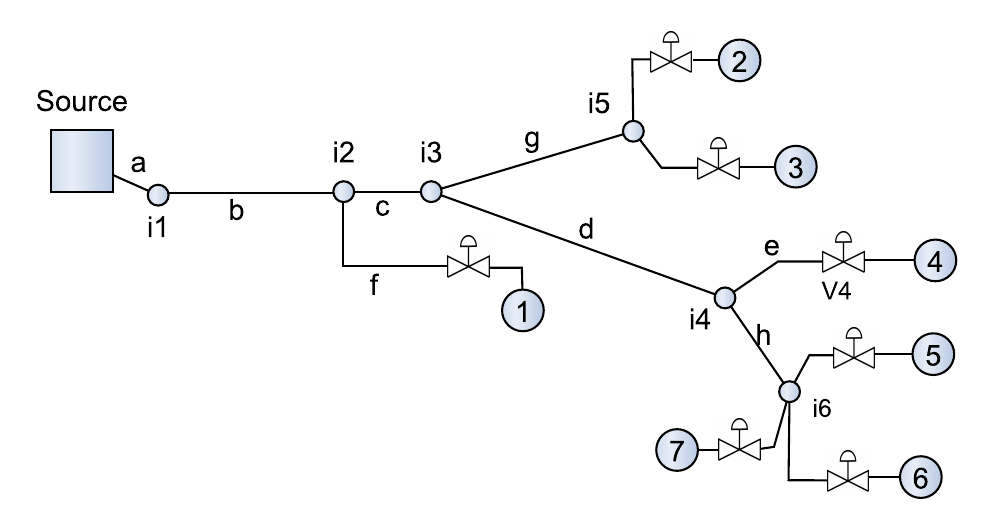}
	\caption{A sample network with a single source and seven demand nodes. }
	\label{fig:prop_1}
	\end{center}
\end{figure}

In this section, we  state  a few results for the system described earlier in Section \ref{sec:sys_description}. 
which are useful in deriving the proofs for the main results in Appendix \ref{sec:results_proof}. The first three results are intuitive and commonly accepted. These are stated as propositions without proofs. The fourth result is presented using arguments that has their basis on the convexity of network flows. The two remaining results are inequalities applicable for real numbers and are  proved using basic results from algebra. 
\begin{prop} \label{prop_1}
In a network as described in Section \ref{sec:sys_description}, partial (or complete) closure of a valve leading to a demand point cannot increase the flow rate into the particular demand point and at the same time, cannot reduce the flow rate to any other demand point. 
\end{prop}
\noindent \emph{Illustrative example:} Consider the partial (or complete) closure of a valve, say $V4$, in the network given in Figure \ref{fig:prop_1}. 
The resistance in the path from source to demand node $4$ increases and therefore, the flow rate in this path has to come down. However, the resistance  to any other demand point cannot increase, and hence, the flow rate towards them cannot reduce. Effectively, only node $4$ receives a reduced flow rate. Additionally, the reduced utilization of links $a$, $b$, $c$ and $d$ caused by the lower flow through $e$ can induce higher flow rates to other demand points in the network.

\begin{figure} [htpb]
    \begin{center}
	\includegraphics[width=0.7\linewidth]{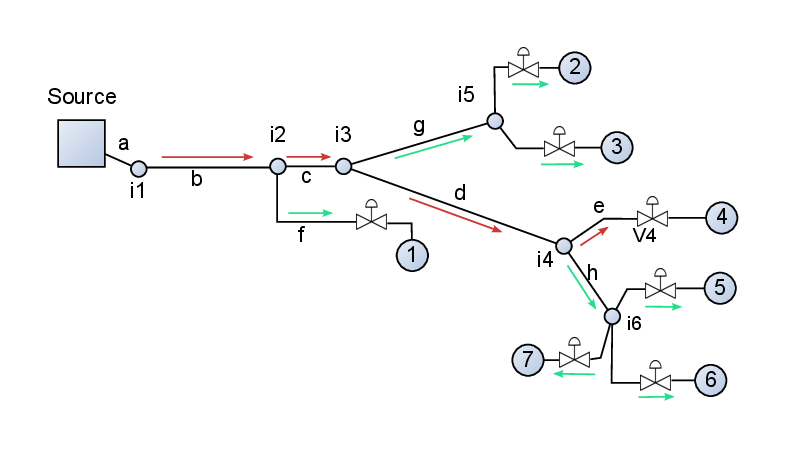}
	\caption{Changes in edge flows with (partial) closure of valve $V4$. A green arrow indicates that the flow rate either increases or remain constant in the corresponding edge. A red arrow indicates that the corresponding flow rates reduces or remain constant.}
	\label{fig:prop_2}
	\end{center}
\end{figure}

\begin{prop} \label{prop_2}
On partial (or complete) closure of a valve leading to a demand point, the reduction in flow rate to the particular demand point cannot be less than the sum of increase in flow rates to other demand points.  
\end{prop}
\noindent \emph{Illustrative example:} 
Figure \ref{fig:prop_2} shows the changes in flow rates with the partial closure of valve $V4$. In edges marked with a red arrow, the flow rate either decreases or remains constant. In edges marked with a green arrow, the flow rate either increases or remains constant.

Closure of a valve can only increase (or keep constant) the net resistance of any network. The total outflow from the source cannot increase on closure of a valve. Consequently, the reduction in flow rate for the demand point with the valve closure ($V4$ here) has to be greater than the sum of the increase for other demand points. For this reason, edges $b$, $c$, and $d$ has an overall reduction in flow as shown in Figure \ref{fig:prop_2}.  

\begin{prop}\label{prop_3}
On partial (or complete) closure of a valve leading to a leaf node, the available head (potential) at every other node in the network does not reduce. 
\end{prop}
\noindent \emph{Illustrative example:} As in Proposition \ref{prop_1}, assume $V4$ is being closed. Consider any node, say $i2$,  on the path from source to $V4$. From Proposition \ref{prop_1}, flow rate in the path reduces on the closure of the valve and hence the head-loss (potential drop) in the links  $a$, and $b$  comes down. As the head at the source remains constant, the available head (potential) at $i2$ increases. 

\begin{figure} [htpb]
    \begin{center}
	\includegraphics[width=0.7\linewidth]{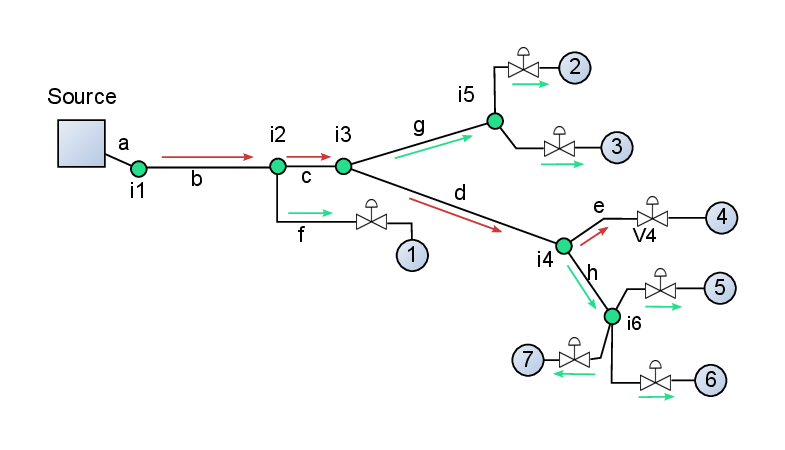}
	\caption{Changes in head at nodes with partial closure of valve $V4$. The green coloration of nodes indicates that the available head increases. }
	\label{fig:prop_3}
	\end{center}
\end{figure}

Now consider another node $i5$ not in the path from source to $V4$. Flow downstream of node $i5$ increases on closure of $V4$ (Proposition \ref{prop_1}). This is possible only if $i5$ is maintained at a higher head than earlier. Similarly, the head (potential) at all nodes in the network increases with the closure of $V4$  as shown in Figure \ref{fig:prop_3}.

\begin{figure}[htpb]
\centering
   \includegraphics[scale=0.4]{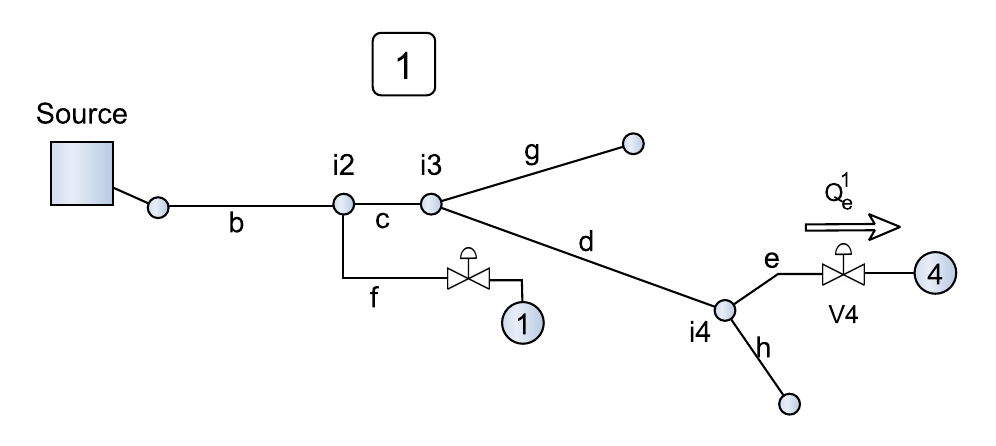}
  \centering
   \includegraphics[scale=0.4]{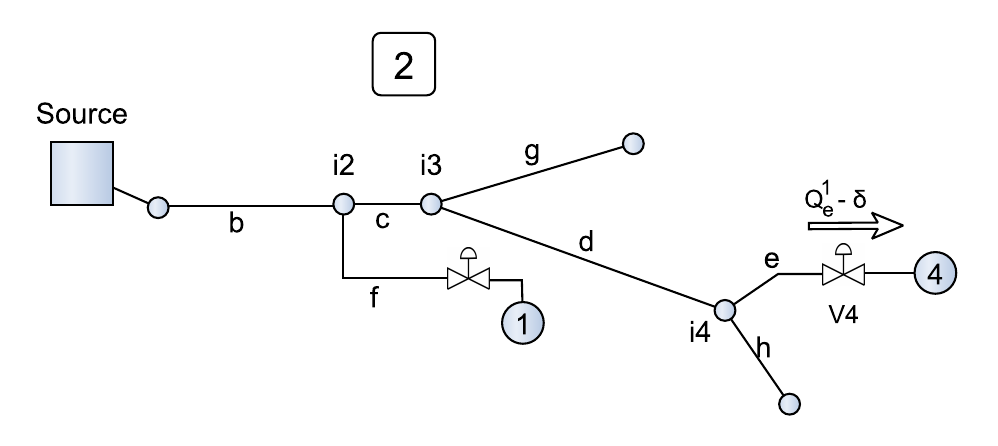}
\centering
 \includegraphics[scale=0.4]{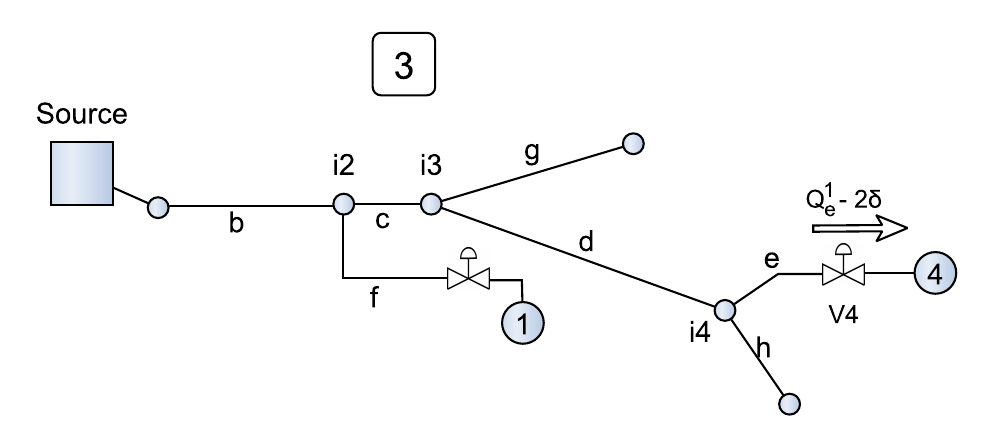}
\caption{Three instances of a skeletonized form of the network given in Figure \ref{fig:prop_1}. In the first instance, the flow rate though edge $e$ is $Q_e^{(1)}$. In the second instance, valve $V4$ is partially closed to reduce the flow through $e$ by an amount $\delta$. In the third instant, the flow through $e$ is further reduced by an amount $\delta$. }
\label{fig:prop_4}
\end{figure}

\begin{prop} \label{prop_4}
Let the partial closure of a valve $V$ reduce the flow rate through it by $\delta$. The  corresponding increase in the sum of flow rates into other demand nodes be $\Delta_1$. Now, on further reducing  the flow rate through $V$ by the same extend $\delta$, let the corresponding increase for other nodes be $\Delta_2$. For a network of the type described in Section \ref{sec:sys_description}, $\Delta_1 \geq \Delta_2$. 
\end{prop}
Consider three instances of a network given in Figure \ref{fig:prop_4}. The networks are obtained after skeletonizing the system shown in Figure \ref{fig:prop_1}. 
As this step involves manipulation of the flow on edge $e$, the path from source to $e$ is kept intact. All branches emerging out of this main line is approximated with an an equivalent single demand node for the sake of convenience. 
All arguments given below are applicable to the original network as well.

Here we use the notation $Q_j^{(i)}$ to denote the flow rate through $j^{th}$ edge in the $i^{th}$ instance. Also, $H_k^{(i)}$ refers to the head at $k^{th}$ node in $i^{th}$ instance. For example, $Q_d^{(2)}$ denotes the flow rate through edge $d$ in $2^{nd}$ instance. The difference in the second instance from the first is that valve $V4$ is partially closed 
to reduce the flow rate through edge $e$ by an amount $\delta$. In the third instance, $V4$ is closed  to reduce the flow rate through $e$ further by an amount $\delta$. That is, $Q_e^{(2)}=Q_e^{(1)}-\delta$ and $Q_e^{(3)}=Q_e^{(1)}-2\delta$. 

The analysis begins with the first link out of the source (edge $b$). Following this, similar arguments are made for for all links in the path from source to 
$e$ in a sequential manner.   

\begin{figure} [htpb]
    \begin{center}
	\includegraphics[scale=0.75]{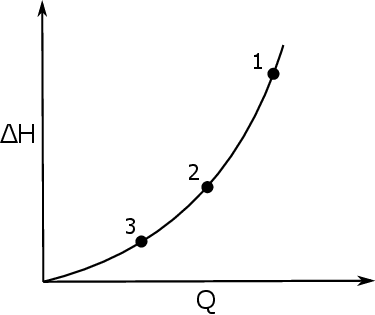}
	\caption{A schematic showing the variation in head loss with flow rate in a pipe. As $\Delta H$ increases super-linearly with $Q$, the function is convex. }
	\label{fig:convex}
	\end{center}
\end{figure}

As explained in Proposition \ref{prop_1}, $Q_b$ reduces monotonically as we move from network instance $1$ to $3$. For the given network, claim in Proposition \ref{prop_4} can be stated as follows: 
\begin{align}
    Q_b^{(1)}-Q_b^{(2)} \leq Q_b^{(2)}-Q_b^{(3)} \label{eq2.0}
\end{align} 
Let us assume the contrary, i.e.
\begin{align}
 Q_b^{(1)}-Q_b^{(2)} > Q_b^{(2)}-Q_b^{(3)}   \label{eq2}
\end{align}
In any line $j$, the rate of change of head-loss with flow rate increases with flow rate as shown by Fig \ref{fig:convex}. 

From the convexity of the relation between $\Delta H$ and $Q$, one may write,
\begin{align}
    \frac{\Delta H^{(1)} - \Delta H^{(2)}}{Q^{(1)} - Q^{(2)}} &>     \frac{\Delta H^{(1)} - \Delta H^{(3)}}{Q^{(1)} - Q^{(3)}} & \forall Q^{(2)} > Q^{(3)} \label{eq:cnvx1}\\
    \frac{\Delta H^{(1)} - \Delta H^{(3)}}{Q^{(1)} - Q^{(3)}} &>     \frac{\Delta H^{(2)} - \Delta H^{(3)}}{Q^{(2)} - Q^{(3)}} & \forall Q^{(1)} > Q^{(2)}   \label{eq:cnvx2} 
\end{align}

From Equations \ref{eq:cnvx1} and \ref{eq:cnvx2}, is is easy to infer that
\begin{align}
    \frac{\Delta H^{(1)} - \Delta H^{(2)}}{Q^{(1)} - Q^{(2)}} &>     \frac{\Delta H^{(2)} - \Delta H^{(3)}}{Q^{(2)} - Q^{(3)}} & \forall \begin{aligned} &\left(Q^{(1)} > Q^{(2)} \right)\\ & \land \left( Q^{(2)} > Q^{(3)}\right)\end{aligned} \label{eq:cnvx3} 
\end{align}

Going back to the flow rates realized in edge $b$ of the network, as $Q_b^{(1)}>Q_b^{(2)}>Q_b^{(3)}$, one may write,
\begin{align}
    \frac{\Delta H_b^{(1)} - \Delta H_b^{(2)}}{Q_b^{(1)} - Q_b^{(2)}} &>     \frac{\Delta H_b^{(2)} - \Delta H_b^{(3)}}{Q_b^{(2)} - Q_b^{(3)}} & \label{eq:cnvx4}
\end{align}
\begin{align}
\text{from (\ref{eq2}) and (\ref{eq:cnvx4})}, ~ H_{i2}^{(2)}-H_{i2}^{(1)} & > H_{i2}^{(3)}-H_{i2}^{(2)} \label{eq6}
\end{align}

With an expression available about the change in the head at the intermediate node $i2$, we observe the flow rate to the first branch emerging out of the main line.  Clearly, it is the head available at node $i2$ that drives the flow through $f$. An increase in the head at $i2$ increases the flow rate  through $f$. However, as evident from Figure \ref{fig:convex}, if the existing flow through the line $f$ is high, the difficulty level of further increasing the flow rate is also high. Therefore, from Equation \ref{eq6}, we can conclude the following.
\begin{align}
 Q_f^{(2)}-Q_f^{(1)} & > Q_f^{(3)}-Q_f^{(2)}   \label{eq7}\end{align}
 From (\ref{eq2}) and (\ref{eq7}), 
 \begin{align}
 Q_c^{(1)}-Q_c^{(2)} & > Q_c^{(2)}-Q_c^{(3)}    \label{eq8}
\end{align}
We made an assumption on edge $b$ in Equation \ref{eq2} and from this, we derived the same relation for edge $c$ in Equation \ref{eq8}. Continuing this argument, we can arrive at similar results for the flow in edge $g$  as we had obtained for the line $f$ in Equation \ref{eq7}. This would result in a relation similar to Equations \ref{eq2} and \ref{eq8}, but this time for edge $d$. We can continue the process until we reach the node form which the edge containing valve $V4$ originates ($i4$ here). Following arguments   similar to the ones made for $f$ and $g$ earlier and applying them to edge $h$ we  arrive at Equation \ref{eq9}. Note that  $g$ and $h$ represent equivalent pipes (resistance) for the actual sub-network at the respective location.  
\begin{align}
 Q_e^{(1)}-Q_e^{(2)} > Q_e^{(2)}-Q_e^{(3)}   \label{eq9}
\end{align}
However, we already know that, 
\begin{align}
 Q_e^{(1)}-Q_e^{(2)} = Q_e^{(2)}-Q_e^{(3)} =\delta  \label{eq10}
\end{align}
Equation \ref{eq9} and \ref{eq10} contradict each other and therefore, the assumption we made earlier, i.e. Equation \ref{eq2} has to be wrong and the relation \ref{eq2.0} has to be correct. This shows that Proposition \ref{prop_4} is valid for the network given in Figure \ref{fig:prop_4}. Also, none of the arguments placed so far is limited to the network given in Figure \ref{fig:prop_4}. We did not assume any specialty for the network (other than the conditions given in Section \ref{sec:sys_description}) and similar arguments can be made for any given network. Hence the validity of Proposition \ref{prop_4} extends to the whole class of networks we address in this paper. 

\begin{prop} \label{prop_5}
For any real numbers $x_1$, $x_2$ and $n$ such that $x_1,~ x_2 \geq 0$ and $n\geq 1$, the following relation holds true.
\begin{align}
    (x_1+x_2)^{1/n} \leq x_1^{1/n}+x_2^{1/n}
    \label{eq12}
\end{align}
\end{prop}
Both sides of Equation \ref{eq12} are positive. Hence, we can equivalently prove the following relation where both sides of the original expression are raised to the power $n$. 
\begin{align}
    \left ( (x_1+x_2)^{1/n} \right )^n & \leq \left ( x_1^{1/n}+x_2^{1/n} \right)^n  \label{eq13}\\
    \text{Left Hand Side(LHS)} &=  x_1+x_2  \label{eq14}\\
    \text{Right Hand Side(RHS)} &=\left ( x_1^{1/n}+x_2^{1/n} \right)^n\\
    &=\left ( x_1^{1/n} \right )^n + \left ( x_2^{1/n} \right )^n \nonumber\\
    &~~~~~ + other~positive~terms \nonumber \\
    & \geq x_1 + x_2  \nonumber\\
    &= LHS \label{eq15}
\end{align}

\begin{prop} \label{prop_6}
For any set of positive real numbers $x_1,~ x_2,~ x_3, \cdots ,x_m ~ $ and $n\geq1$, the following bound holds true:
\begin{align}
    \frac{x_1^{1/n}+x_2^{1/n}+x_3^{1/n}+\cdots +x_m^{1/n}}{\left (x_1+x_2+x_3+ \cdots +x_m \right )^{1/n}} \leq m^{1-1/n} \label{eq17}
\end{align}
Further, the inequality (\ref{eq17}) is strict unless all $x$'s are equal. 
\end{prop}
Beckenbach and Bellman \cite{beckenbachinequalities} define \textit{Mean of order t} of values in an array $\mathbf{y}$ with weight $\mathbf{\alpha}$ 
as follows:
\begin{align*}
    M_t(y,\alpha) \equiv \left ( \displaystyle \sum_{i=1}^{m} \alpha_i y_i^t \right )^{1/t}
\end{align*}
The authors show that for positive $\mathbf{y}$ and any real $t$, $M$ is a non-decreasing function of $t$. \begin{align}
    \frac{d M_t(y,\alpha)}{dt}\geq 0 \label{eq18}
\end{align}
Further, unless all values of $\mathbf{y}$ are equal, $M$ is strictly increasing with $t$. 

Let there be a weight vector $\mathbf{\alpha}$ with its entries equal to $1/m$ where $m$ is the number of entries in $\mathbf{y}$. From Equation \ref{eq18}, one may infer that following holds true for any $n\geq 1$.
\begin{align}
    \left ( \displaystyle \sum_{i=1}^{m} \frac{1}{m} y_i^n \right )^{1/n} \geq \left ( \displaystyle \sum_{i=1}^{m} \frac{1}{m} y_i \right )
\end{align}
\noindent Replacing $y_i$ with $x_i^{1/n}$, 
\begin{align*}
    \left ( \displaystyle \sum_{i=1}^{m} \frac{1}{m} x_i \right )^{1/n} & \geq \left ( \displaystyle \sum_{i=1}^{m} \frac{1}{m} x_i^{1/n} \right )\\
    \frac{\left (x_1+x_2
    + \cdots +x_m \right )^{1/n}}{m^{1/n}} & \geq \frac{x_1^{1/n}+x_2^{1/n} 
    +\cdots +x_m^{1/n}}{m}\\
 m^{1-1/n} & \geq \frac{x_1^{1/n}+x_2^{1/n}
 +\cdots +x_m^{1/n}}{\left (x_1+x_2
 + \cdots +x_m \right )^{1/n}} \\
 QED
\end{align*}

\newpage

\section{An upper bound on relative time of operation} \label{sec:results_proof}
A mathematical definition for the upper bound on the relative time of operation is ($\mathbf{R}^*$) is given by Equation \ref{eq:proof_1}.

\begin{align}
    \mathbf{R}^*=\displaystyle \sup_{\substack{N \in \mathbb{N}, K \in \mathbb{K}_N,\\ D \in \mathbb{D}_N}} \left [\frac{\displaystyle \min_{t} \sum_{(d,i)\in \mathbb{U}_d}t_{d,i}~\left (N,K,D\right) }{\displaystyle \min_{t} \sum_{(cv,i)\in \mathbb{U}_{cv}}t_{cv,i} ~ \left (N,K,D\right )}\right] \text{where,} \label{eq:proof_1}
\end{align} 

\begin{tabular}{ll}
$\mathbb{N}$     & is the set of network topologies\\
&falling under the description in Section \ref{sec:sys_description}\\
$\mathbb{K}_N$     & is the set of all pipe resistances for topology $N$  \\
$\mathbb{D}_N$     & is the set of all network demands  for topology $N$\\
$\mathbb{U}_{d}$     & is the set of all valve configurations for a system\\
&with discrete valves\\
$\mathbb{U}_{cv}$     & is the set of all valve configurations for a system\\
&with continuous valves\\
$t_i$     & is the time for which state $i$ is active\\ 
\end{tabular}

This section presents the derivation of an upper bound for $\mathbf{R}^*$, w.r.t. a general system described in Section \ref{sec:sys_description}.  The bound is derived in three major steps. Deriving $\mathbf{R}^*$ with a  complete schedule of continuous control valves is a difficult task. However, it is easier to find it if the continuous control valves are operated only in a single configuration.  The first step shows that the supremum for $\mathbf{R}$ given by the latter case can be as high as the same obtained in the former. In spite of this simplification, there still  exist infinite types of networks that one needs to search through. In the second step, it is shown that the supremum for $\mathbf{R}$, i.e. $\mathbf{R}^*$, is given by a specific class of network. The last step is the derivation of $\mathbf{R}^*$ for this particular network configuration. Each step ends with one of the results given in Section \ref{sec:Results}.

\subsection{$\mathbf{R}^*$ is attained while operating continuous control valves in a single configuration}\label{sec:proof_result1}
To deliver a given quantity of water to each demand point, a network with continuous control valves may be operated in different configurations (states) for their respective time span. These configurations and their corresponding active time form the  schedule of operation. In the following paragraphs, it is shown that, in the current problem, one needs to consider only those schedules in which continuous control valves are operated in a single configuration. 

For this, Expression \ref{eq:proof_1} is re-written as follows:
\begin{align}
    \mathbf{R}^*=\displaystyle \max_{\substack{N \in \mathbb{N}, K \in \mathbb{K}_N}} \left [\frac{\displaystyle \min_{t} \sum_{(d,i)\in \mathbb{U}_d}t_{d,i} ~ \left(N,K,D_{cv}\right) }{\displaystyle \min_{t} \sum_{\substack{(cv,i)\in \mathbb{U}_{cv}\\t_{cv,i}>0}}t_{cv,i} ~ \left(N,K\right)}\right]  \label{eq:proof_2}
\end{align}

\noindent {where $D_{cv}$ is the demand met when the system is operated using continuous control valves for times $t_{cv,i}\left(\forall\left(cv,i\right)\right)$ .} There are two main differences between   (\ref{eq:proof_2}) and  (\ref{eq:proof_1}). (i) The  denominator now contains only the states of continuous control valves that are active. 
This makes no change to the value of the denominator because the time intervals corresponding to inactive states are anyways zero. (ii) The search space  includes  all network topologies and resistances but excludes explicit specifications of   demands {in the denominator}. 
This too does not affect the optimal solution as the entire space of network topologies, resistances (including pipe resistance in a scheduled operation of continuous control valves), {and active times of the states $\left(t_{cv,i}\right)$ span the complete set of network demands as well. It may be noted that the demand met using continuous control valves (denominator) becomes a constraint while identifying the minimum time schedule for operating the system with discrete valves (numerator). }

\begin{figure}[htpb]
    \centering
    \begin{subfigure}[t]{0.5\linewidth}
        \centering
	\includegraphics[scale = 0.5]{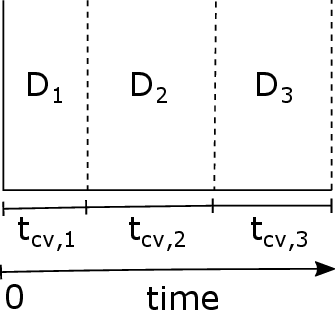}
        \caption{}
        \label{fig_2_1}
    \end{subfigure}%
    \begin{subfigure}[t]{0.5\linewidth}
        \centering
	\includegraphics[scale = 0.5]{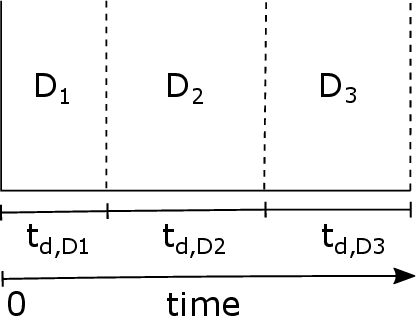}
        \caption{}
         \label{fig_2_2}
    \end{subfigure}
    \caption{Schedule for operation using continuous control valves (a) and discrete valves (b). In (a), $D_1$, $D_2$, etc. denote the demand satisfied by separate configurations of valves. In (b), $t_{d,D1}$, $t_{d,D2}$ etc. denote the minimum time required to meet the same demands $D_1$, $D_2$, etc. using discrete valves.}
\end{figure}

In the operation of continuous control valves, there may be multiple states active. Let the time intervals corresponding to each of these be $t_{cv,1}$, $t_{cv,2}$, $t_{cv,3}$ etc. Let the quantity of water supplied  each of these configurations be denoted as  $D_1$, $D_2$, $D_3$ etc, where the $D_i$'s are demand vectors (for all the demand nodes)
.  The total demand $D_{cv}$ would be met in time $t_{cv}$ and let $t_{cv}= t_{cv,1}+t_{cv,2}+t_{cv,3}+\cdots = 1$. For any $t_{cv} \neq 1$, one could scale it up/down to 1 after applying the same scaling to the demand as well. The first three states of such a schedule is represented by Figure \ref{fig_2_1}.

In the next case, the same system is operated, but with discrete valves, to meet demand $D_1$ in the minimum time possible. Let this be achieved in time $t_{d,D1}$ as shown in Figure \ref{fig_2_2}. It has to be noted that, the system may be operated in multiple configurations within this time interval. Following this, the system is further operated for  time $t_{d,D2}$ to meet the demand $D_2$ in the minimum possible time. Likewise for $D_3$ and so on. The  total time $t_d$ is defined as $t_d=t_{d,D1}+ t_{d,D2}+t_{d,D3} + \cdots$. In time $t_d$,  demands $D_1$, $D_2$ etc.  are met, which would sum up to the total demand $D$. 
$t_d$ may be greater than the minimum time required to meet the overall demand because here we solve separate scheduling problems to minimize $t_{d,D1}$, $t_{d,D2}$, etc. rather than a single one to minimize $t_d$

The ratio of total operational times, $\mathbf{R} \equiv \frac{t_{d}}{t_{cv}}$. 
Further, $\mathbf{R_1} \equiv \frac{t_{d,D1}}{t_{cv,1}}$ denotes the ratio of operational times, had the demand been equal to $D_1$. Accordingly, the following are defined: 
$\mathbf{R_2} \equiv \frac{t_{d,D2}}{t_{cv,2}}, ~~ \mathbf{R_3} \equiv \frac{t_{d,D3}}{t_{cv,3}},\ldots$ 

From expression \ref{eq:proof_2} one may write
\begin{align}
    \mathbf{R}^* & \leq \displaystyle \max_{\substack{N \in \mathbb{N},\\ K \in \mathbb{K}_N}} 
    \frac{\left [\displaystyle \min_{t} \left (\begin{aligned} &t_{d,D1}\left(N,K\right)\\ &~~~+t_{d,D2}\left(N,K\right)\\ &~~~~~~+t_{d,D3}\left(N,K\right)\\ &~~~~~~~~~ + \cdots 1 \end{aligned}\right )\right]}{1} 
    \label{eq:proof_3}\\
    & =\displaystyle \max_{\substack{N \in \mathbb{N},\\ K \in \mathbb{K}_N}} 
    \frac{\left [\displaystyle \min_{t} \left (\begin{aligned} &\mathbf{R_1}\left(N,K\right)t_{cv,1}\left(N,K\right) \\ &~~~+ \mathbf{R_2}\left(N,K\right)t_{cv,2}\left(N,K\right)\\ &~~~~~~+\mathbf{R_3}\left(N,K\right)t_{cv,3}\left(N,K\right)\\ &~~~~~~~~~+ \cdots \end{aligned}\right )\right]}{1} 
    \label{eq:proof_4}\\
    & = \begin{aligned}\displaystyle \max_{\substack{N \in \mathbb{N},\\ K \in \mathbb{K}_N}} \left [\displaystyle \max \left( \left \{\begin{aligned} &\mathbf{R_1}\left(N,K\right),\\ &~~~\mathbf{R_2}\left(N,K\right),\\ &~~~~~~\mathbf{R_3}\left(N,K\right), \cdots\end{aligned} \right \} \right)  \right] & \\~ \because t_{cv,1}+ t_{cv,2}+  \cdots  =1 &\end{aligned}\label{eq:proof_5}
\end{align}




Expression \ref{eq:proof_5} implies that the ratio of operational times for meeting the complete demand is no more than the highest of the corresponding ratio obtained for separate demands $D_1$, $D_2$, etc. Each of these individual demands denotes the water delivered by a single configuration of the network using continuous control valves. That is, the value of $\mathbf{R}$ obtained from a scheduled operation of continuous control valves cannot be greater than the maximum of the same obtained from a single configuration. This simplifies the problem substantially. Rather than searching for the supremum of $\mathbf{R}$ with a scheduled operation of valves, one can restrict the search to cases with only a single configuration of continuous control valves.  

Further, any value  $\mathbf{R}_i$ obtained with a single configuration of continuous control valves lie in the feasible space for $\mathbf{R}$. Hence, the inequality given by Equation\ref{eq:proof_5} can be considered a strict equality:

\begin{align}
    \mathbf{R}^*& = \displaystyle \max_{\substack{N \in \mathbb{N}, K \in \mathbb{K}_N, \\D \in \mathbb{D'}_N}} \left [\frac{\displaystyle \min_{t} \sum_{(d,i)\in \mathbb{U}_d}t_{d,i}\left(N,K,D\right)}{1} \right]  \label{eq:proof_7}
\end{align}
    
\noindent where $\mathbb{D'}_N$ refers to the set of demands that can be met by continuous control valves in a single state, operated for one unit of time. 

Now it remains to identify the particular configuration and demands for which $\mathbf{R}$ reaches its maximum. This is carried out in the following sections.


\subsection{Networks achieving maximum $\mathbf{R}$} \label{sec:proof_result2}
As shown in Expression \ref{eq:proof_7}, the inner optimization problem tries to minimize the time taken by discrete valves to meet the demand. Now suppose instead of solving the minimization problem, we propose the following general operational schedule for discrete valves (denoted $\mathcal{S}$). 

\noindent \textbf{Schedule} $\mathcal{S}$: \textit{Open all valves initially. Continue supply until the demand is met for any demand node. Close the valve corresponding to the demand point and continue supplying to other locations. Repeat the procedure and stop once the demand is met for all nodes.}

The heuristic schedule $\mathcal{S}$ need not be optimal and hence Inequality \ref{eq:proof_8} follows.
\begin{align}
    \mathbf{R}^*& \leq \displaystyle \max_{\substack{N \in \mathbb{N}, K \in \mathbb{K}_N, \\D \in \mathbb{D'}_N}} \left [\displaystyle \sum_{(d,i)\in \mathcal{S}}t_{d,i}\left(N,K,D\right)  \right]
    \label{eq:proof_8}
\end{align}

\begin{figure} [htpb]
    \begin{center}
	\includegraphics[width=0.7\linewidth]{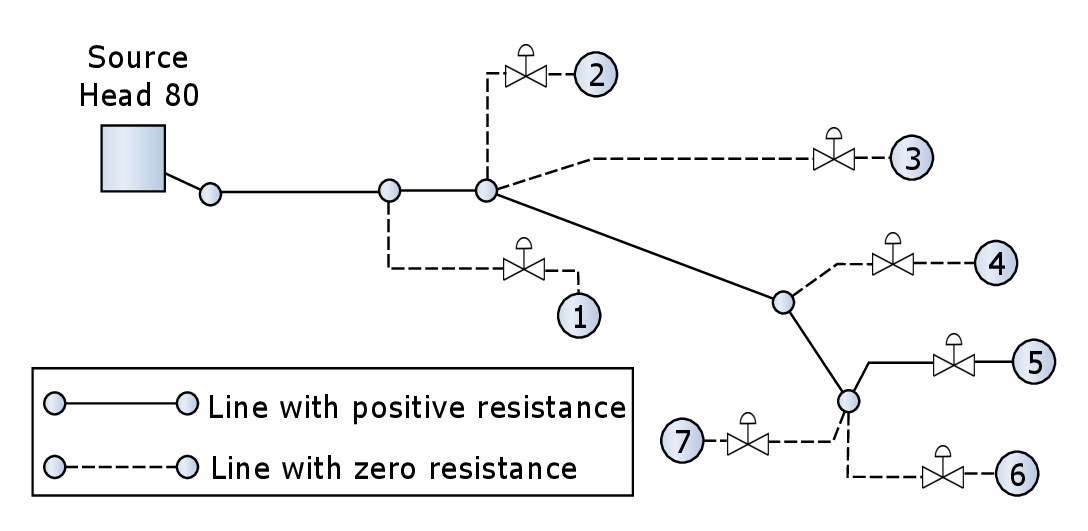}
	\caption{Network configuration that maximizes $R$. Here only the line from source to node $5$ have positive resistances when fully open. All branches emerging out of this main line have zero resistance. $\mathcal{C}$ denotes the class of such networks.}
	\label{fig:topology_optimal_R}
	\end{center}
\end{figure}

The task at hand is to find the network topology and resistances which may lead to maximizing the time taken by discrete valves operated as per schedule $\mathcal{S}$. We claim that this is realized in networks satisfying a specific condition on line resistances. Here, the main line -  connecting the source to one of the leaf nodes - would have positive resistance. All other demand points would be connected directly to this main line with a pipe of zero resistance when completely open. The corresponding configuration for Figure \ref{fig:sample_network1} is given in Figure \ref{fig:topology_optimal_R}. Let $\mathcal{C}$ denote the class of networks having this configuration.  


To prove that the maximum value of $\mathbf{R}$ is attained for the particular configuration, three cases are discussed below. In each, the topology and the  amount of water supplied to the demand points are the same. 

\subsubsection{Case 1, Continuous control valves}
Consider the network shown in Figure \ref{fig:sample_network1} and valves  initially be of continuous type. Let the system be operated for a unit of time in any configuration. The same amount of water supplied here has to be repeated in the following cases as well. 

\subsubsection{Case 2, A schedule for the system with discrete valves} \label{sec:schedule_system_discrete}
\begin{figure} [htpb]
    \begin{center}
	\includegraphics[scale=0.7]{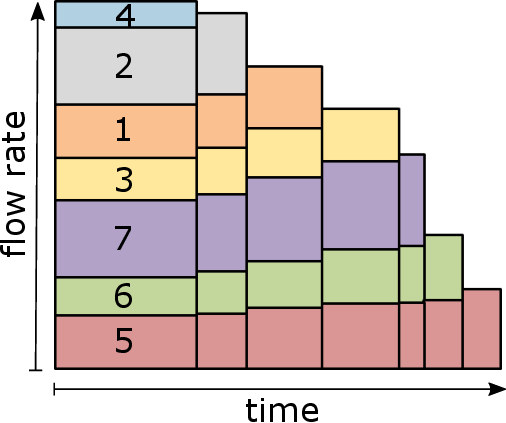}
	\caption{Operational policy for system with discrete valves. Here each vertical bar corresponds to a particular network configuration or state. The width of the bar denotes the time for which the state is active and the height of the bar denotes the flow rate in the state. Each cell represents the flow rate received by individual demand points. The network continues operation in a state until the demand for any node is satisfied, after which, the corresponding valve is closed. The numbering of demand nodes and the flow rates are representative. }
	\label{fig:schedule_discrete}
	\end{center}
\end{figure}

As a second case, consider again the same system, but now with the valves changed to be of discrete (ON/OFF) type and the same water has to be supplied. The operational strategy is to supply water following schedule $\mathcal{S}$. A schematic showing the flow rates in this operational policy is given in Figure \ref{fig:schedule_discrete}. Here, each horizontal bar denotes a configuration of the system and its width corresponds to the time for which the state is active. The height of each cell indicates the flow rate received by each demand point. Different colors are used to indicate different demand points as mentioned in the first bar. The total operational time ($t_d$) for the schedule is the time for which the final valve stays open. Let $\mathcal{V}$ be this final valve.


\subsubsection{Case 3, A modified network}
In the third case, consider a network with the same topology as Case 2  described earlier in Section \ref{sec:schedule_system_discrete}, but belonging to class $\mathcal{C}$. That is the configuration that was earlier hypothesized to have the highest value for $\mathbf{R}$. The path from source to $\mathcal{V}$ (the valve that was open throughout in Case 2) shall have the same resistance as that of Case 2. All other demand points shall be connected directly to the main line at the corresponding locations using a pipe of zero resistance. The following paragraphs show that operating this network using discrete valves requires more (or equal) time compared with the Case 2.

\begin{figure} [htpb]
    \begin{center}
	\includegraphics[scale=0.5]{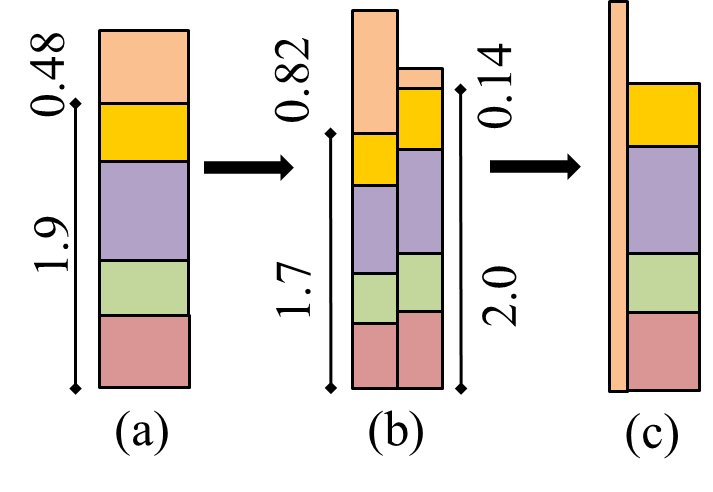}
	\caption{(a) Flow rates in one state of the network. Here each color represent a demand point. The height of the cell denotes the flow rate received by the particular node  and the width of the cell corresponds to the time for which the state is active. (b) A combination of two states that are active for the same time as (a). The flow rates received by the demand point closest to the source (denoted by the rectangle on top) is varied in the two states by changing the setting of the respective valve. However the total amount of water supplied to the node remains the same as that of (a). (c) Increasing the change in valve setting finally leads to the limiting case where the node closest to the source receives all water in one of the states. The total active time of the states in (c) and the water received by the node closest to the source are same as that of (a) and (b). The numbers on the y-axis are representative values for flow rates. }
	\label{fig:state_flows}
	\end{center}
\end{figure}

To start with, let us assume that the network has continuous control valves in pipes leading to all demand points.
The flow rates obtained in every active network configuration of Case 2 can be achieved here with appropriate valve settings. W.l.o.g. let Figure \ref{fig:state_flows}(a) represent the flow rates and active time of one active state. Here each color represents the flow to  a particular node. The height and width of the rectangle correspond to the magnitude of flow rate and time for which the state is active, respectively. The area of each rectangle, therefore, corresponds to the quantity of water delivered to each demand point by the particular state. 

Let the rectangle on top correspond to the demand point closest to the source. It is possible to increase or reduce the flow rate to this node by adjusting the control valve. In the next step, the configuration (a) is replaced with with two states operating for the same duration, as shown in (b). In the two new states, the flow rate to the node closest to the source (node 1) is increased and decreased respectively such that total water supplied to the particular node (node 1) is maintained the same. The flow rate into other nodes (nodes 3, 7, 6, and 5) would be lower than (a) in the first state of (b) and larger than (a) in the second state of (b). This follows from Proposition \ref{prop_1}. It can further be inferred from Proposition \ref{prop_4} that the total water delivered to other nodes is lower (or equal) in (b) as compared to (a). 

Continuing the process (of increasing and decreasing flow rates in the two states), one arrives at two states as shown in (c). Here, in the first state, the valve leading to the node closest to the source is fully open and  only this node receives the water as the resistance to this node from the main line is zero. Thereafter, in the second state, other nodes receive water. Also, the total amount of water supplied to demand points would be less than  the water supplied  in (a) except for the node closest to the source (node 1), for which it would remain the same. It may also be noted that the operational time for (a), (b), and (c) are all the same.

The procedure may be repeated for the second state of (c) starting with the node that is now closest to the source. Finally, what remains would be states having only one demand point receiving supply at a time. However, these states would require more (or equal) time as compared to (a) to supply the same amount of water. Further, the same arguments can be extended to other states active in Case 2. The only state that remains unchanged would be the last state, i.e. with only valve $\mathcal{V}$ open. The pipe leading to valve $\mathcal{V}$ has the same resistance in both Case 2 \& 3. 
Finally, the resultant schedule would have water supplied to one demand point at a time. Two important observations about this final schedule are:  
\begin{itemize}
    \item The new schedule of supplying one demand point at a time can provide only a lower  (or equal) amount of water in the given amount of time. In other words, the time required to meet the total demand would at least as much for which Case 2 was active.
    \item This new schedule, with only one valve open at a time, may be implemented on any network belonging to class $\mathcal{C}$ with only ON/OFF valves. In fact, this is the only schedule implementable on networks belonging to $\mathcal{C}$ when equipped with ON/OFF valves. As all except one demand node is connected to the main line with zero resistance, opening a demand point implies it withdraws the entire water present in the main line.  Any node connected further downstream does not receive any water at this time.
\end{itemize} 

Therefore, the modified network Case 3, when equipped with discrete valves, would need more time to deliver water than Case 2. 

For any given instance of Case 2, one can have a corresponding instance of Case 3 which requires a larger (or equal) amount of time  to deliver the water. Therefore, it can be concluded that the class of networks $\mathcal{C}$ - with only the path to one demand node having positive resistance - is the configuration that attains the highest value for \textbf{R} for a given topology. This simplifies the problem significantly. The next task is to find an upper bound on \textbf{R} from among the class of networks belonging to $\mathcal{C}$.  

There is one caveat: A single network topology can have multiple configurations belonging to $\mathcal{C}$. To be more precise, there is one configuration corresponding to each demand node being the part of the main line. Depending on the water to be delivered and the network resistances, Schedule $\mathcal{S}$ can have any valve to close last. For each of these instances, there exists a separate network configuration in $\mathcal{C}$.  

With the results obtained from the paragraphs above, we may re-write the expression for $\mathbf{R}^*$
\begin{align}
    \mathbf{R}^*& \leq \displaystyle \max_{\substack{N \in \mathbb{N}, K \in \mathbb{K}_{N}, \\D \in \mathbb{D'}_N}} \left [\displaystyle \sum_{(d,i)\in \mathcal{S}}t_{d,i}\left(N,K,D\right) \right]
    \label{eq:proof_8_1}\\
    & = \displaystyle \max_{\substack{N \in \mathbb{N}, K \in \mathbb{K}_{N,\mathcal{C}}, \\D \in \mathbb{D'}}} \left [\displaystyle \sum_{(d,i)\in \mathcal{S}}t_{d,i}\left(N,K,D\right) \right] 
    \label{eq:proof_9}\\
    &= \displaystyle \max_{\substack{N \in \mathbb{N}
    }} \left [ \displaystyle \max_{j} \left ( \left \{ \displaystyle \max_{\substack{K \in \mathbb{K}_{N,\mathcal{C}_j},\\ D \in \mathbb{D}'_N}
    }\sum_{(d,i)\in \mathcal{S}}t_{d,i}\left(N,K,D\right) \right\} \right ) \right] ~ \text{ }  \label{eq:proof_10}
\end{align}

\noindent where, $\mathcal{C}_j$ is a subset of class $\mathcal{C}$ that has demand node `$j$' connected to the network with a line of positive resistance and all other leaf nodes connected to the main line with a line of zero resistance. $\mathbb{K}_{N,\mathcal{C}_j}$ is the set of network resistances feasible for networks belonging to class $\mathcal{C}_j$.


Earlier, Expression \ref{eq:proof_8}, was written as an inequality rather than an equality as the schedule $\mathcal{S}$ proposed for discrete valves need not have been an optimal schedule. However, when it comes to networks belonging to class $\mathcal{C}$, supplying water to one demand point at a time (schedule $\mathcal{S}$) is the only option available. Hence, this also has to be the optimal schedule. This allows us to write the Expression \ref{eq:proof_10} as an equality.
\begin{align}
    \mathbf{R}^*= \displaystyle \max_{\substack{N \in \mathbb{N},
    }} \left [ \displaystyle \max_{j} \left ( \left \{ \displaystyle \max_{\substack{K \in \mathbb{K}_{N,\mathcal{C}_j},\\ D \in \mathbb{D}'_N }}\sum_{(d,i)\in \mathcal{S}}t_{d,i}\left(N,K,D\right) \right\} \right ) \right]   \label{eq:proof_11}
\end{align}

In the following section \ref{sec:An_expression_for_R*}, an explicit expression is derived  for the  supremum of the innermost optimization problem of Expression \ref{eq:proof_11}.

\subsection{An expression for $\mathbf{R}^*$} \label{sec:An_expression_for_R*}
In the previous sections, it was shown that the highest value of $\mathbf{R}$ is attained by a particular network configuration $\mathcal{C}$. In this section, a value for this expression is derived.

\subsubsection{The class of networks $\mathcal{C}$}
\begin{figure} [htpb]
    \begin{center}
	\includegraphics[width=0.7\linewidth]{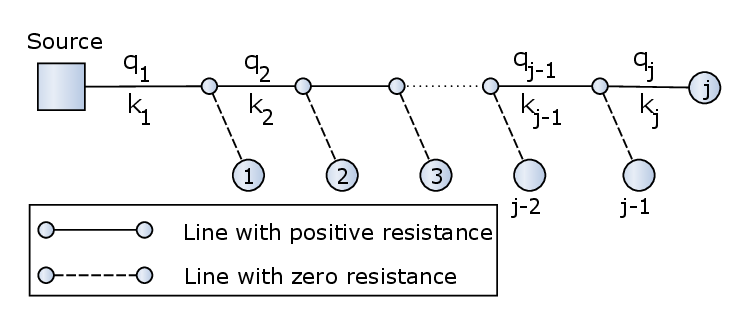}
	\caption{A general network belonging to class $\mathcal{C}$. Here one origin - consumer path has positive resistance. All other demand points are connected to the main line with an edge of zero resistance. }
	\label{fig:final_topology}
	\end{center}
\end{figure}

Consider a network with one origin-consumer path of positive resistance and all other demand points connected to this \textit{main line} with zero resistance. Any such network can be considered equivalent to the structure shown in Figure \ref{fig:final_topology}. Even if there are multiple demand points that are connected to the \textit{main line} at the same point, these can be lumped into a single node for which, the demand is equal to the sum of demands of all nodes it replaces. It is safe to ignore the network topology between the replaced nodes and the main line as these are of zero resistance.  The expression $\mathbf{R}^*$, can therefore, be derived with respect to networks of the type shown in Figure \ref{fig:final_topology}. 

\subsubsection{$\mathbf{R}^*$ for the class of networks $\mathcal{C}$} \label{sec:R*}
Consider the network given in Figure \ref{fig:final_topology}. The network has one main line of positive resistance and branches of zero resistance emerging out of it. Let $k_1$, $k_2$, $k_3$, $\cdots$, $k_j$ be the resistances of the pipes in the main line connecting source to node $j$. Also, let  $q_1$, $q_2$, $q_3$, $\cdots$, $q_j$ be the flow rates through them while operating with with continuous control valves. For any non-negative set of values for $k$'s, it is always possible to have any non-negative set of values for $q$'s (such that $q_1 \geq q_2 \geq... q_j $) by suitably adjusting the continuous control valves on the zero-resistance edges and the total head available at the source. However, specifying both $k$'s and $q$'s uniquely identifies the state  of the network and there is no more degree of freedom left.

This configuration is active for one unit of time. Later, when the system is operated with discrete valves, the active states would have only one node supplied at a time. The time for which node $i$ is supplied water $\left(t_{d,i} \right)$ is given by the following relations. 
\begin{align*}
t_{d,1}&=\frac{q_1-q_2}{\left (\frac{\Delta h}{k_1}\right )^{1/n}}\\
t_{d,2}&=\frac{q_2-q_3}{\left (\frac{\Delta h}{k_1+k_2}\right )^{1/n}}\\
t_{d,j-1}&=\frac{q_{j}-q_{j-1}}{\left (\frac{\Delta h}{k_1+k_2+ \cdots k_{j-1}}\right )^{1/n}}\\
t_j&=\frac{q_{j}}{\left (\frac{\Delta h}{k_1+k_2+ \cdots k_{d,j}}\right )^{1/n}}
\end{align*}

The total time required to supply water using discrete valves $\left (t_{d} \right)$ is given by:

\begin{align}
    t_{d} &=\displaystyle \sum_{\left (d,i \right) \in \mathcal{S}} t_{d,i}  \\
    &=\frac{\left [ \begin{aligned} &q_1 \left (k_1^{1/n} \right )\\ 
    &~~~ +q_2 \left ( (k_1+k_2)^{1/n} - (k_1)^{1/n} \right ) + \cdots \\ 
    &~~~~~~ + q_{j} \left ( \begin{aligned} &(k_1+k_2+ \cdots + k_m)^{1/n}\\ &~~~ - (k_1+k_2+ \cdots + k_{j-1})^{1/n} \end{aligned} \right) \end{aligned} \right ]}{{\Delta h}^{1/n}} \label{eq:8_7}
\end{align}
    Substituting for $h$ and using the algebraic inequality given in Equation \ref{eq12}
\begin{align}     
    ~~&\leq \frac{q_1k_1^{1/n}+ q_2k_2^{1/n}+ \cdots + q_mk_m^{1/n}}{\left ( k_1q_1^n+ k_2q_2^n+ k_3q_3^n+\cdots +k_j q_j^n\right )^{1/n}} \label{eq:8_8}  
\end{align}
    Replacing $k_iq_i^{n}$ by a new non-negative variable $x_i$
\begin{align}    
    ~~& = \frac{x_1^{1/n}+ x_2^{1/n}+x_3^{1/n}+\cdots +x_j^{1/n}}{\left (x_1+x_2+x_3+ \cdots +x_j \right )^{1/n}} \label{eq:8_9} 
\end{align}

The objective is to find an upper bound for $t_{d}$. This is equivalent to finding an upper bound for the expression given in \ref{eq:8_9} after imposing no constraints on $x$ except for the non-negativity constraints\footnote{Balance of node flows require $q_1\geq q_2\geq \cdots \geq q_j$ and the derivation of Equation \ref{eq:8_8} from Equation \ref{eq:8_7} assumes  $k_1 \ll k_2 \ll \cdots \ll k_j$. However $x$ may still take any non-negative value.}. Proposition \ref{prop_6} provides a tight upper bound for this as given in Inequality \ref{eq:8_9_1}. 
\begin{align}
\frac{x_1^{1/n}+ x_2^{1/n}+x_3^{1/n}+\cdots +x_j^{1/n}}{\left (x_1+x_2+x_3+ \cdots +x_j \right )^{1/n}} \leq j^{1-1/n} \label{eq:8_9_1}
\end{align}

As the system with continuous control valves was operated for one unit time, the supremum of $t_{d}$ also equals the ratio $\mathbf{R}^*$. 
\begin{align}
    \mathbf{R}^* = \displaystyle \max_{\substack{N \in \mathbb{N}}} \left [ \displaystyle \max_{j} \left ( j^{1-1/n} \right ) \right] \text{where,}  \label{eq:proof_12}
\end{align}

\noindent $j$ is the depth of the main path. Clearly, the higher the value of $j$, higher the bound on $\mathbf{R}$. As we seek an upper bound, the value of $j$ for which the RHS of Expression \ref{eq:proof_12} reaches its maximum is taken. That is, when $j$ is the maximum depth of the network in consideration. 

Therefore, the supremum of $\mathbf{R}$ is given as follows:
\begin{align}
    \mathbf{R}^* =  m^{1-1/n}   \label{eq:proof_13}
\end{align}

\noindent where $m$ is the maximum depth of the network in consideration and $n$ is the  coefficient in Equation \ref{eq1} relating head loss to flow rate in a pipe.

\end{document}